\documentclass[]{aa}
\usepackage{psfig}                          
\def\lya{Ly$\alpha$ }   
\newcommand{\be}{\begin{equation}} 
\newcommand{\en}{\end{equation}}

\def\zabs{$z_{\rm abs}$} 
 
\def\lya{Ly$\alpha$ } 
 
\def\h2{H$_2$}

\def\siiv{Si~{\sc iv}~} 
\def\siiva{Si~{\sc iv}$\lambda$~1393~} 
\def\siivb{Si~{\sc iv}$\lambda$1402~}

\def\kms{km~s$^{-1}$} 
\def\dela{$\Delta\alpha/\alpha$~}
 \def\chisq{$\chi^{2}$~} 
\begin{document} 
\title{{Probing the time-variation of the fine-structure constant: Results
based on Si IV doublets from a UVES sample} 
\thanks{ 
Based on observations collected 
at the European Southern Observatory (ESO), under the Large Programme  
ID No. 166.A-0106 with UVES on the 8.2m Kuyen 
telescope operated at the Paranal Observatory, Chile}}  
\titlerunning{fine-structure constant} 
\author{Hum Chand\inst{1}, Patrick Petitjean\inst{2,3}, Raghunathan Srianand\inst{1}  
\& Bastien Aracil\inst{2,4}} 
\institute{$^1$IUCAA, Post Bag 4, Ganeshkhind, Pune 411 007, India\\ 
$^2$Institut d'Astrophysique de Paris -- CNRS, 98bis Boulevard 
Arago, F-75014 Paris, France\\ 
$^3$LERMA, Observatoire de Paris, 61 Rue de l'Observatoire, 
   F-75014 Paris, France\\ 
$^4$Department of Astronomy, University of Massachusetts, 
710 North Pleasant Street, Amherst, MA 01003-9305, USA\\ 
} 
\date{Received date/ Accepted date} 
\offprints{H. Chand \\~\email{hcverma@iucaa.ernet.in}} 
\abstract{ 
We report 
a new constraint on the variation of the fine-structure constant 
based on the analysis of 15 Si~{\sc iv} doublets selected from a 
ESO-UVES sample. We find \dela ~=~${ (+0.15\pm0.43)\times10^{-5}}$ over  
a redshift range of ${ 1.59\le z\le 2.92}$ which is consistent with no  
variation in $\alpha$.  
This result represents a factor of three improvement on the constraint on \dela  
based on \siiv doublets compared to  
the published results in the literature. 
The alkali doublet method used here avoids the implicit assumptions used
 in the many-multiplet method that chemical and ionization inhomogeneities 
 are negligible and isotopic abundances are close 
to the terrestrial value.
\keywords{ 
{\em Quasars:} absorption lines -- 
{\em cosmology:} observations 
} 
} 
\maketitle 
\section{Introduction} 
Some of the modern theories of fundamental physics, in particular SUSY GUT and 
Super-string theory, motivate experimental searches of  
possible variations in the fine-structure constant. Such theories  
require the existence of extra `compactified'  
spatial dimensions and allow for the  
cosmological evolution of their scale size. As a result, these theories 
naturally predict the cosmological  
variation of fundamental constants in  
a 4-dimensional sub-space (Uzan 2003 and reference therein).\par 
In the framework of the standard Big-bang model, 
quasar spectra can be used as an important tool 
to test  the variation of the fine-structure constant, $\alpha=e^2/\hbar c$, by allowing one to measure 
its value  at different redshifts. 
Bahcall, Sargent \& Schmidt (1967) were the first 
to use the absorption lines of alkali-doublets seen 
in QSO spectra to constrain the variation of this quantity. 
Their analysis provided   
 ${\Delta\alpha/\alpha} \equiv(\alpha_{z}-\alpha_{0})/\alpha_{0}= (-2\pm 5)\times 10^{-2}$ 
 at a redshift $z$~$\sim$~1.95. Here $\alpha_{0}$   
 refers to the value of the fine-structure constant on Earth  
 and  $\alpha_{\rm z}$ to its value at redshift $z$. 
 Since then several authors have used  the alkali-doublet 
  method (AD method) to constrain the variation of $\alpha$ 
 (Wolfe, Brown \& Roberts 1976; Levshakov 1994; 
 Potekhin \& Varshalovich 1994; Cowie \& Songaila, 1995; 
 Varshalovich, Panchuk \& Ivanchik, 1996; 
 Varshalovich, Potekhin \& Ivanchik 2000; 
 Martinez, Vladilo \& Bonifacio, 2003).  
 The method is based on the fact that the separation between 
 energy levels caused by  
 fine-structure interactions is proportional to $\alpha^4$ with the leading  
 term of energy level being proportional to $\alpha^2$. As a result, to a very  
 high  accuracy,   
the relative separation of a fine-structure doublet,  
$(\lambda_{2}-\lambda_{1})/\lambda=\Delta\lambda/\lambda$,  
 will be proportional to $\alpha^2$. 
 Here $\lambda_{1}$ and $\lambda_{2}$ are, respectively, the rest wavelength 
 corresponding to transition $^2$S$_{1/2}\rightarrow^2$P$_{3/2}$ and 
  $^2$S$_{1/2}\rightarrow^2$P$_{1/2}$ of the alkali-doublet and 
  ~$\lambda$ is the average value of $\lambda_{1}$ and $\lambda_{2}$. 
 Varshalovich, Potekhin \& Ivanchik (2000) give 
the following relation between  \dela   
and the values of ($\Delta\lambda/\lambda$) at redshifts 0 and $z$: 
 \begin{equation} 
  \frac{\Delta\alpha}{\alpha_{ }}=\frac {cr}{2}~\bigg{[}\frac{(\Delta\lambda/\lambda)_{z}}{(\Delta\lambda/\lambda)_{0}}-1\bigg{]}, 
 \end{equation}   
 where ``$cr$''($\approx 1$) represents the higher order relativistic correction.\par
Actually, the dependence of rest wavelengths to the
 variation  of $\alpha$ 
is  parameterized  using the fitting function  
 given by Dzuba et al. (1999a) 
\begin{equation} 
\omega=\omega_{0}+q_{1}x+q_{2}y 
\end{equation} 
Here $\omega_{0}$ and $\omega$ are, respectively, the {vacuum} wave  
number (in  units of cm$^{-1}$) measured in the laboratory  
and in the absorption system at  redshift $z$.
$x$ and $y$ are the dimensionless numbers 
defined as $ x=(\alpha_{z}/\alpha_{0})^{2}-1$ and 
$ y=(\alpha_{z}/\alpha_{0})^{4}-1 $. The  sensitivity  
coefficients $q_{1}$ and $q_{2}$ are obtained using  
many-body relativistic calculations (see Dzuba et al. 1999a). 
 For a given doublet and $\Delta\alpha/\alpha <<1$, Murphy et al. (2001)  
have shown that Eq. (2) reduces to Eq. (1) with,  
  \[ cr \approx \frac{\delta q_{1} +\delta q_{2}}{\delta q_{1} +2\delta q_{2}} \] 
 The value of ``$cr$'' for Si~{\sc iv} is 0.8914 
 when using the $q$ coefficients  as given in Table~\ref{tabat} (see below). 
\par\noindent
The AD method works with emission as well as absorption lines.  
 However emission lines are usually broad as compared to 
 absorption lines. 
 As a result, the constraints obtained from emission lines are not as precise as 
those derived from  
 absorption lines. Bahcall et al. (2004) have recently found 
 $\Delta\alpha/\alpha$~=~${ (0.7\pm1.4)\times10^{-4}}$ using O~{\sc iii} 
 emission lines from QSOs.  
 The most stringent constraint from alkali-doublet absorption lines  has been  
 obtained by Murphy et al. (2001),  
 $\Delta\alpha/\alpha$~=~${ (-0.5\pm1.3)\times10^{-5}}$,  
 by analyzing a KECK/HIRES sample of 21 Si~{\sc iv} doublets  
 observed along 8 QSO sight lines. 
\par\noindent
The generalization of this method known as the many-multiplet (MM)  
method (Dzuba et al. 1999b) 
makes use of a combination of transitions from different species. 
The sensitivity coefficients $q_{1}$ and $q_{2}$ 
of heavier elements are found to be an order of 
magnitude higher than those of lighter elements. 
 As a result the MM method gives an order of magnitude better  
 precision in the measurement of $\Delta\alpha/\alpha$.
 Application of MM method to KECK/HIRES  
 data resulted in the measurement of  
 $\Delta\alpha/\alpha$~=~${ (-0.57\pm0.10)\times10^{-5}}$  
 over the redshift range $0.2\le z\le 3.7$ (Murphy et al. 2003). 
 However our recent investigation (Srianand et al. 2004 and Chand et al. 2004) 
 using very high quality UVES data and well defined selection criteria  
 resulted instead in a null detection of \dela 
 ($\Delta\alpha/\alpha$~=~${ (-0.06\pm0.06)\times10^{-5}}$) 
 over the redshift range $0.4\le z\le 2.3$.
\par\noindent
\begin{table} 
\caption{ List of Si~{\sc iv} doublets in our sample} 
\begin{tabular}{lllll} 
\hline\hline 
\multicolumn{1}{c}{QSO}   &{$z_{em}$}   & {$z_{abs}$} &  Comments\\
\hline 
HE~1341$-$1020    & 2.135  & 1.915             &  \\ 
               &        & 2.147             & saturated \\ 
Q~0122$-$380      & 2.190  & 1.906             &     \\ 
               &        & 1.969             &     \\ 
               &        & 1.973             &     \\ 
               &        & 1.975             &     \\ 
PKS~0237$-$23     & 2.222  & 1.597$^{*}$       &    \\ 
HE~0001$-$2340    & 2.263  & 2.183              &  \\ 
Q~0109$-$3518     &        & 2.045           & contaminated \\ 
HE~2217$-$2818    & 2.414  & 1.965           &  weak \& blend\\ 
               &        & 2.186           &  blend\\ 
Q~0329$-$385      & 2.435  & 2.251           & contaminated \\ 
HE~1158$-$1843    & 2.449  & 2.266           & blend \\

HE~1347$-$2457    & 2.611  & 2.329            &      \\ 
Q~0453$-$423      & 2.658  & 2.276            & contaminated    \\ 
               &        & 2.502            & blend   \\ 
PKS~0329$-$255    & 2.703  & 2.328            & unstable  \\ 
               &        & 2.454           &   \\ 
               &        & 2.455           &   \\ 
Q~0002$-$422      & 2.767  & 2.167$^{*}$     &     \\ 
               &        & 2.301              & contaminated  \\ 
               &        & 2.464             &     \\ 
HE~0151$-$4326    & 2.789  & 2.451             &    \\ 
               &        & 2.493             &    \\ 
HE~2347$-$4342    & 2.871  & 2.735             & contaminated   \\ 
HE~0940$-$1050    & 3.084  & 2.667             &  contaminated \\ 
               &        & 2.828            &  \\ 
               &        & 2.830            &  unstable\\ 
               &        & 2.916            &  blend\\ 
PKS~2126$-$158    & 3.280  & 2.727             & contaminated  \\  
               &        & 2.768             & contaminated \\ 
               &        & 2.907             & contaminated \\ 
Q~0420$-$388      & 3.117  & 3.087             & broad \\ 
\hline 
\\ 
\multicolumn{4}{l}{`$^{*}$' \siiv system below \lya emission }\\ 
\multicolumn{4}{l}{~~~~~but included in our sample (see text).}\\ 
\multicolumn{4}{l}{``contaminated'' \siiv doublet with inconsistent}\\ 
\multicolumn{4}{l}{~~~~~profiles due to contamination of absorption} \\ 
\multicolumn{4}{l}{~~~~~lines from other systems.} \\ 
\multicolumn{4}{l}{``blend'' the majority of the components} \\ 
\multicolumn{4}{l}{~~~~~have separations less than the} \\
\multicolumn{4}{l}{~~~~~individual $b$ values.} \\
\multicolumn{4}{l}{``unstable'' \siiv doublet with unstable Voigt}\\
\multicolumn{4}{l}{~~~~~profile fit.}\\ 
\label{tablist} 
\end{tabular} 
\end{table} 
However,
these results based on the MM method hinge on two assumptions:(i)
ionization and chemical homogeneity and (ii) isotopic abundances of
 Mg~{\sc ii} close to the terrestrial value. Even though these are reasonable
 assumptions one cannot completely rule out systematic biases 
induced by them in the analysis, especially when one is looking
for very small effects. On the contrary one can completely avoid the assumption
of homogeneity in the case of the AD method because, by construction, the two lines
of the doublet must have the same profile (see also Quast et al. 2004).
Also the effect of isotopic shifts is negligible in the case of 
Si~{\sc iv} doublets (see Section 3.4 of Murphy et al. 2001). 
 Therefore it is important to increase the precision of \dela measurements 
based on the AD method. This can be achieved by 
 (a) increasing the S/N ratio and spectral resolution of the data used; 
 (b) increasing the sample size.  
 The S/N ratio of the data used by Murphy et al. (2001) is in the range  
15-40 per pixel and the spectral resolution is $R\sim34000$. 
 In this paper our  motivation is to  improve the \dela~ measurements by  
 using the alkali doublets detected in our UVES data of higher S/N and 
resolution. 
Si~{\sc iv} is used instead of C~{\sc iv} because wavelengths are  
better known for Si~{\sc iv} (Griesman \& Kling 2000; Petitjean \& Aracil 2004a) 
and $q$ coefficients are larger (see Table 2). 
\par
\noindent 
 The organization of the paper is as follows. 
 In Section 2 we briefly describe our data sample and analysis. 
 The importance of selection criteria is discussed in Section 3 and 
 discussion of individual systems are given in Section 4.
The results and overall discussion are presented in Section 5.
 \section{Data sample and analysis} 
\subsection{Data sample}  
The data used in this study have been obtained with the 
Ultra-violet and Visible Echelle Spectrograph (UVES) mounted on the ESO  
Kuyen 8.2~m telescope at the Paranal observatory for the ESO-VLT   
Large Programme ``Cosmic evolution of the intergalactic medium" 
(PI Jacqueline Bergeron). 
This data set corresponds to a homogeneous sample of 18 
QSO lines of sight with a Lyman-$\alpha$ redshift  range of 
1.7 to 3.2. The detailed quantitative description of data calibration 
are presented in Aracil et al. (2003) and  
Chand et al. (2004, here after Paper I).  
Briefly, the data is reduced using the UVES pipeline.  
Addition of individual exposures is performed by a sliding window  
and weighting the signal by the errors in each pixel. The final S/N ratio  
is about 60-80 per pixel and the median resolution  
$R$~$\sim$~45000. The continuum is fitted using an automated continuum fitting  
procedure (Aracil et al. 2003).  
\par 
The \siiv systems detected in our data set are listed in Table~\ref{tablist}. 
There are 31 \siiv systems  
redshifted beyond the \lya emission line from the quasar.  
In addition, two systems, marked with an asterisk (*) in Table~1,  
falling in the Lyman-$\alpha$ forest have well defined narrow components 
and have therefore been incorporated in the sample.\par 
Among the 31 systems (beyond the \lya emission), 9 systems are not considered in the 
analysis because they are contaminated  
by other metal lines and/or  atmospheric absorption.
They are noted as ``contaminated''  
in column 4 of Table~\ref{tablist}.    
In addition, 4 other systems are rejected for the following reasons.  
One is completely saturated   
(``saturated'' in Table~\ref{tablist}) and one is a  
very broad system  (``broad'').
The profile of this broad system is spread over 350 \kms, and has 
a few bad pixels in the central part of the \siivb line.
The other two systems (marked with ``unstable'' in column 4 of Table~\ref{tablist}) 
were rejected during the analysis as we found that the component 
structure of their best-fit is not stable (as discussed in Section 3.2).  
Thus from a total of 33 systems, we exclude 13 systems (9 contaminated,
 one saturated, one very broad and 2 systems having large uncertainties in 
the component structure) and  are left with 
20 systems. \par
\begin{figure} 
\psfig{figure=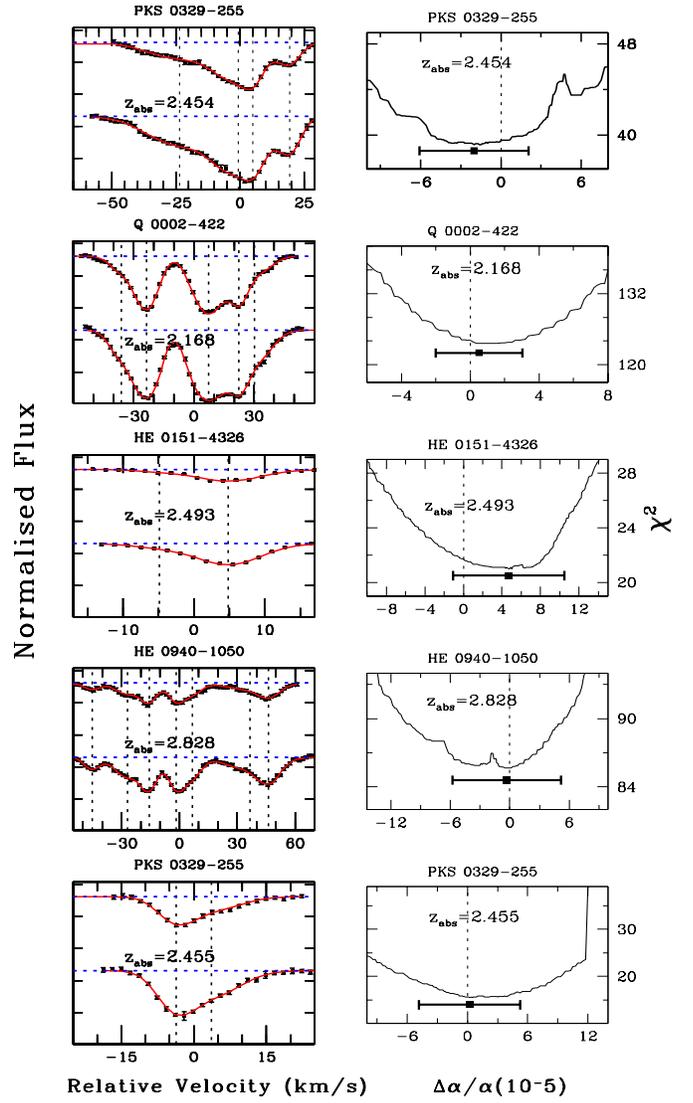,height=15.cm,width=8.8cm,bbllx=54bp,bblly=172bp,bburx=338bp,bbury=706bp,clip=true,angle=0} 
\caption[]{ 
Left panels show on a velocity scale 
Si~{\sc iv} doublet data points with error-bars  
together with the best Voigt-profile fit for \dela$=0$ over plotted
as a solid curve. 
The dotted vertical lines mark the position of components.  
Right panels show the variation of $\chi^2$ 
 as a function of \dela for the systems in the corresponding left panel. 
 Dark rectangles with error bar indicate the measured values of  
 \dela with one sigma error-bar obtained using $\chi^2_{min}+1$ 
 statistics. Name of QSOs and \zabs~ are stated explicitly.} 
\label{fig1fit} 
\end{figure} 
\begin{figure*} 
\psfig{figure=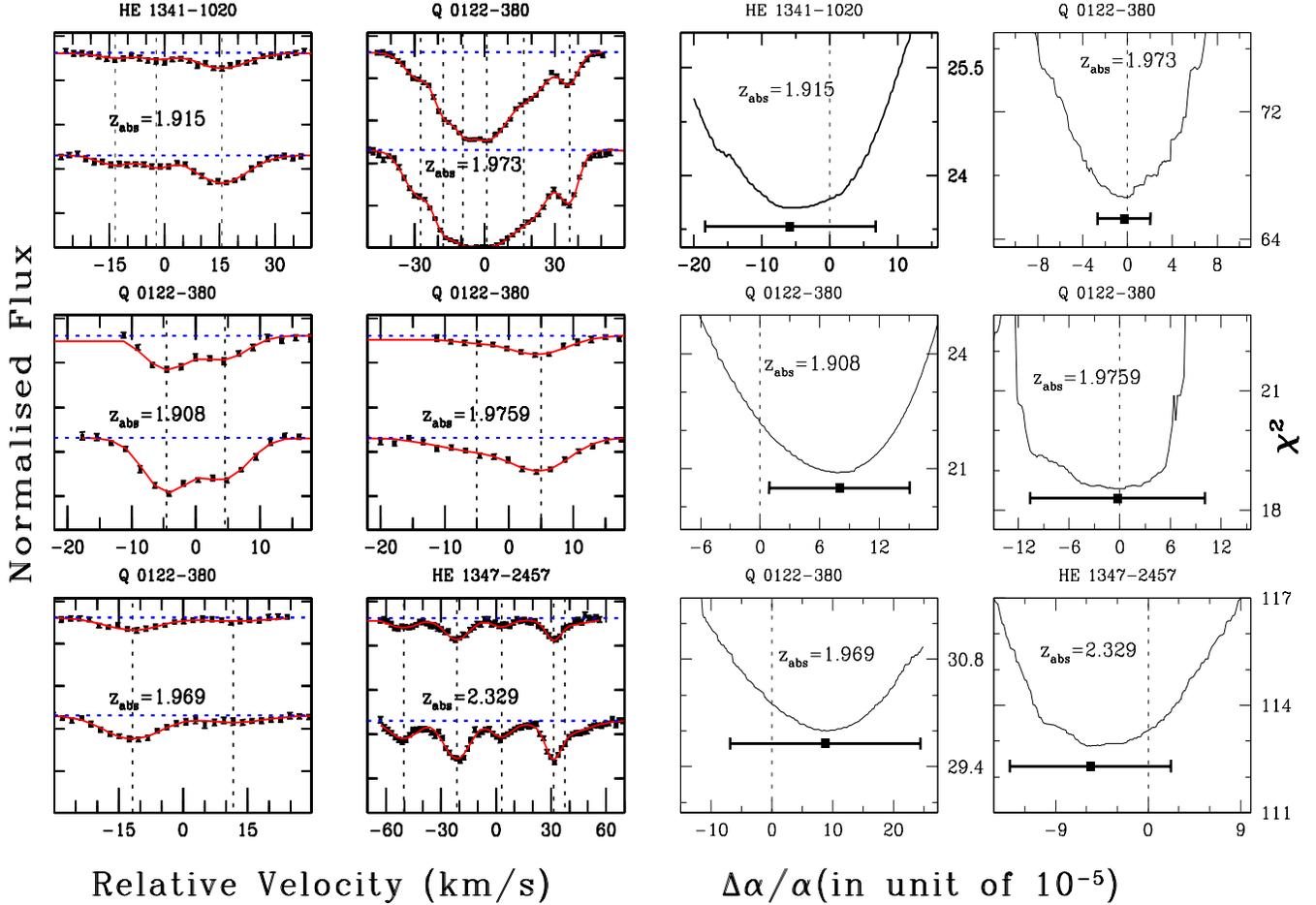,height=13.2cm,width=17.9cm,bbllx=60bp,bblly=370bp,bburx=585bp,bbury=714bp,clip=true,angle=0} 
\caption[]{ 
The first and second panels (from left) show on a velocity scale
Si~{\sc iv} doublet data points with error-bars together with 
the best Voigt-profile fit for \dela$=0$ over plotted as a solid curve. 
 The dotted vertical lines mark the position of components.  
 The third and forth panels show respectively the variation of $\chi^2$ 
 as a function of \dela for the systems in the first and second panels. 
 Dark rectangles with error bar indicate the measured values of  
 \dela with the one sigma error-bar obtained using $\chi^2_{min}+1$ 
 statistics. Name of QSOs and \zabs~ are also stated.
} 
\label{fig2fit} 
\end{figure*}
We have shown in Paper I (using detailed simulations) that 
it is better to avoid weak, or heavily saturated, or strongly blended  
absorption lines in order to obtain a better accuracy on \dela~  
measurements. Indeed, such systems can induce false alarm detection of non-zero \dela. 
\par\noindent
\begin{table} 
\caption{ {\bf Atomic data of Si~{\sc iv} doublet used in our analysis:}} 
{\tiny 
\begin{tabular}{lrrrrr} 
\hline\hline 
\\ 
\multicolumn {1}{c}{Ion} &\multicolumn {1}{c}{$\lambda^{a}$(\AA)} &\multicolumn {1}{c}{$\omega_{0}(cm^{-1})$} &\multicolumn {1}{c}{$q_{1}^{b}$} &\multicolumn {1}{c}{$q_{2}$}&\multicolumn {1}{c} {$f^{c}$}  \\ 
       &               &               & {$(cm^{-1})$} &{$(cm^{-1})$}&\\ 
\hline \\  
Si~{\sc iv}  &1393.76018(4)   &71748.355(2)    &$766$      &$48$   &0.5140 \\ 
Si~{\sc iv}  &1402.77291(4)   &71287.376(2)    &$362$      &$-8$    &0.2553  \\ 
\\ 
\hline 
\\ 
\multicolumn {6}{l} {(a) Griesmann U. \& Kling R. (2000)} \\ 
\multicolumn {6}{l} {(b) Dzuba et al. (1999a) } \\ 
\multicolumn {6}{l} {(c) Martin \& Zalubas (1983) and Kelly (1987)} \\ 
\end{tabular} 
\label{tabat} 
} 
\end{table} 
For a typical S/N ratio of 70 and a median $b$ Doppler parameter of 9 \kms~
as seen in our sample, 
we define a lower limit for $N$(Si~{\sc iv}) of $1.55\times10^{12}$ cm$^{-2}$
so that both lines of the doublets are detected at more than 5$\sigma$ level. 
As in Paper I, we define a multi-component system to be 
unblended if the majority of its components have separations 
larger than the individual $b$ values. 
We apply these criteria after the  
 Voigt profile decomposition of the doublets (as described in the following Section).  
 Based on the best fitted parameters, we  
 decide whether a given system satisfies our selection criteria or not.
Out of 20 systems for which we have performed Voigt profile
fitting, 4 systems are blended and  one system is both 
blended and weak.
We mark these system respectively by `blend' and `weak'
in column 4 of Table~\ref{tablist}.  As a result we are finally left with
15 \siiv doublets that are useful for \dela~ measurements.
The procedure used for the \dela measurement is described 
in the next section.   
\subsection{Analysis} 
We first carry out a Voigt profile fit for each system assuming \dela$=0$. 
The rest wavelengths for the \siiv doublet as well as the other atomic  
parameter used in the fits are summarized in Table \ref{tabat}.  
\begin{figure}
\psfig{figure=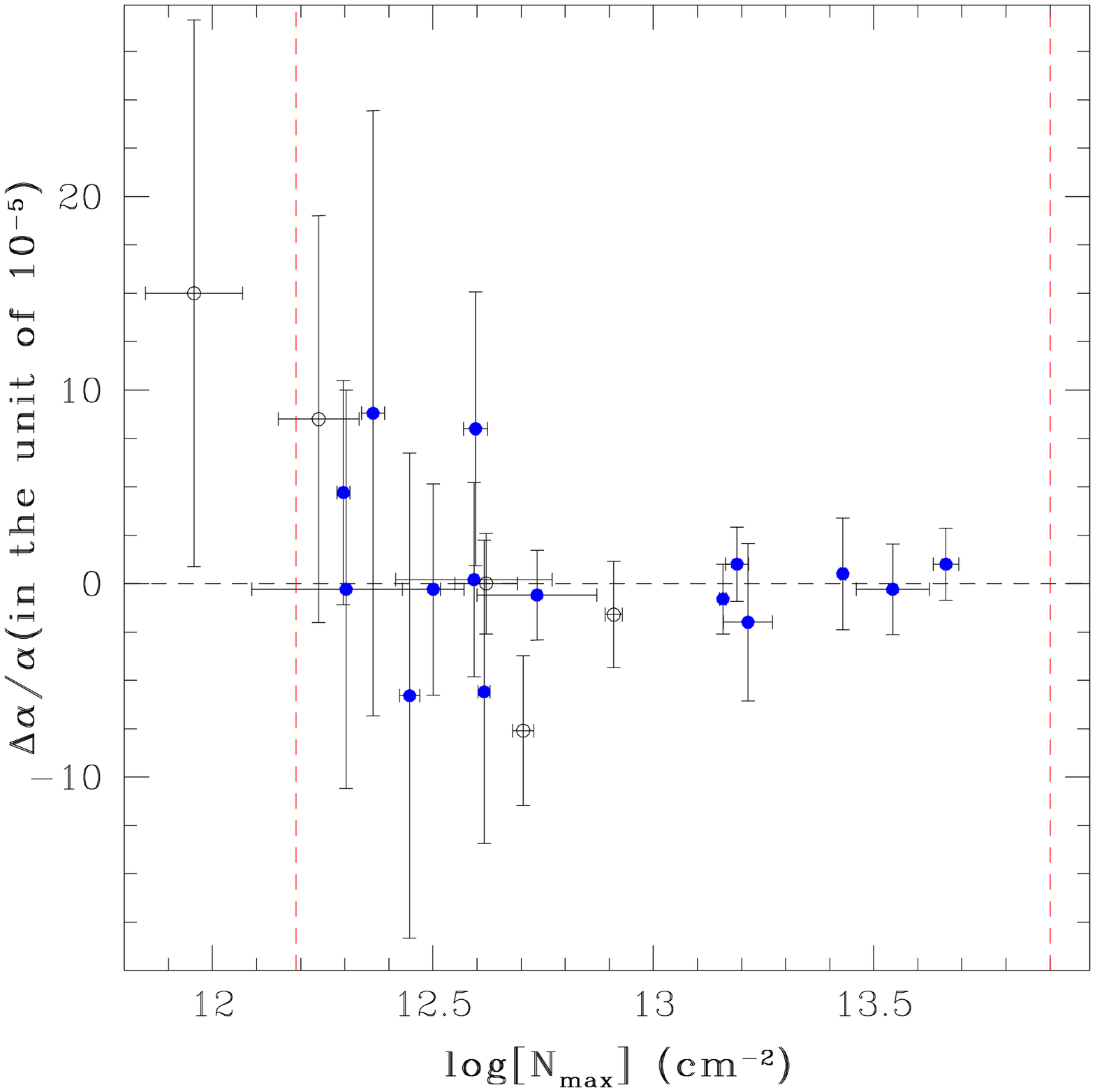,height=8.cm,width=8.cm,angle=00} 
\caption[]{
The figure shows \dela measured 
from the 20 Si~{\sc iv} systems in our sample versus the column density of the 
strongest component in the system. Open circles are for systems that we
define as "blended" (they do not pass our selection criteria) and filled 
circles are for systems that pass this selection criteria. 
It is apparent that errors are larger for weaker systems and for blended systems
as expected from the simulations presented in Paper I.
The vertical dashed lines refer to the lower and upper cut  
off in column density to avoid very weak and heavily  
saturated systems. 
}
\label{figcri} 
\end{figure} 
\begin{figure*} 
\psfig{figure=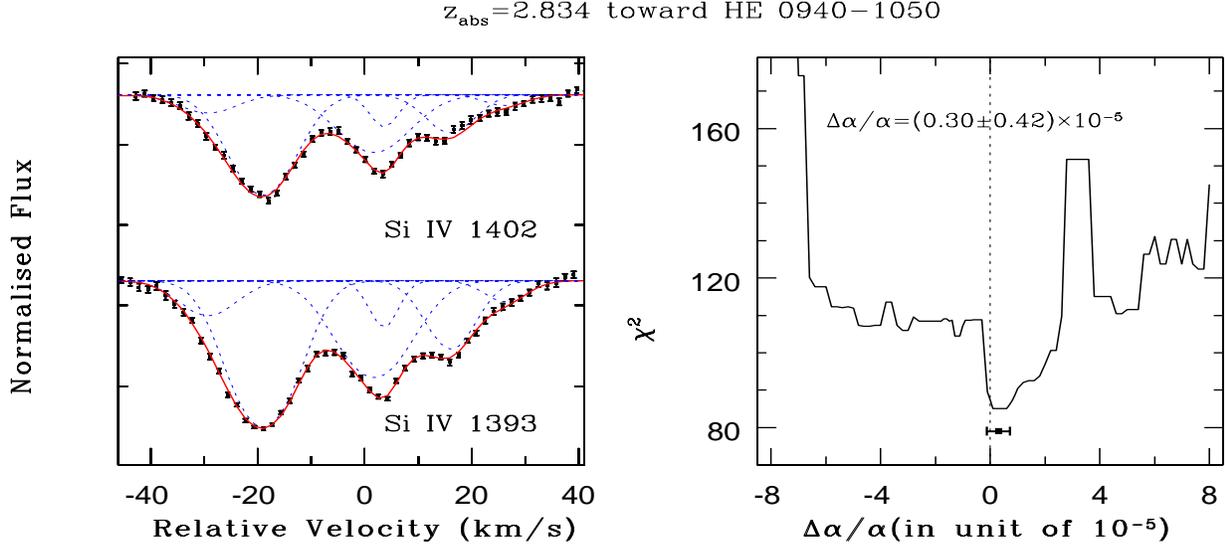,height=7.5cm,width=18.cm,bbllx=0bp,bblly=427bp,bburx=590bp,bbury=718bp,clip=true,angle=0} 
\caption[]{
The left panel shows the Voigt profile best-fit to the system (solid line) along 
 with sub-components profile (dashed lines). The right panel  
 shows the variation of the $ \chi^{2}$ as a function of \dela. The $ \chi^{2}$ 
 curve is median smoothed over a smoothing scale of $1\times10^{-5}$.  
 The large fluctuation in the curve show that the minima is local and 
as a result the system is not considered in the final result. Note that
the instability of the fit mainly comes from the poorly constrained structure
of the central feature at $\sim$5~km~s$^{-1}$.}
\label{fig2p83} 
\end{figure*} 
The Voigt profile fit is carried out by simultaneously varying   
the column density, $N$, Doppler parameter, $b$, and redshift, $z$, for 
each component till the reduced $\chi^2$ of the fit is $\sim 1$. 
This gives us the required number of components to be used in the fit 
and an initial guess of $N$, $b$ and $z$ for each component.\par 
To constrain the variation of $\alpha$, we use  
the analytical expression of the wave number ($\omega$) as a function of  
$\alpha_{z}/\alpha_{0}$ given by Dzuba et al. (1999a), 
\begin{equation} 
\omega=\omega_{0}+q_{1}\bigg{[}\bigg{(}\frac{\alpha_{z}}{\alpha_{0}}\bigg{)}^{2}-1\bigg{]}+q_{2}\bigg{[}\bigg{(}\frac{\alpha_{z}}{\alpha_{0}}\bigg{)}^{4}-1\bigg{]}. 
\end{equation} 
The sensitivity coefficients $q_{1}$ and $q_{2}$, laboratory wave numbers
($\omega_{0}$) and rest wavelengths ($\lambda$)  are listed in Table~\ref{tabat}.
The oscillator strengths ($f$) used in the analysis are given in the last column.\par
We then consider a change in $\Delta\alpha/\alpha$ and fit the system 
using the new rest wavelengths as given in Eq.(3). We vary \dela~ from  
$-20.0\times10^{-5}$ to $20.0\times 10^{-5}$ in steps of $0.1\times10^{-5}$ 
and minimize $\chi^{2}$ for each value \dela by varying N, b, and z.  
The value of $\Delta\alpha/\alpha$ at which $\chi^2$ is minimum ($\chi^2_{\rm min}$)  
is accepted as the measured value of the \dela  from the system, 
provided the reduced $\chi^2$ of the fit is also $\sim 1$. 
 The n$\sigma$ error is obtain from value of \dela~ where $\Delta\chi^{2} =
\chi^{2}-\chi^2_{\rm min}=n^2$, assuming the error on data are
normally distributed (Press et al. 2000, p. 691, Theorem D and Appendix A).
To be on the conservative side we use as 1$\sigma$ error-bar, 
the larger of the two values of  \dela derived using $\chi^{2}-\chi^2_{\rm min}$~=~1
from  the left and right side of $\chi^2_{\rm min}$.
More details about the validation of our fitting procedure using
the simulated data set can be found in Paper I.\par
 In principle we can explicitly use \dela as  one of the  fitting
   parameters (like $N$, $b$ and $z$) in our \chisq
   minimization. However in that case we will have no way of ensuring
   whether the best fitted value of \dela is a true value or an
   artifact  of local minima due to inconsistent fitting of the line profiles.
 Therefore to ensure that the \chisq minima with respect to our crucial
 parameter \dela is not a local minimum, we have preferred to use the 
\chisq versus \dela curve method rather than a single minimization
that would simultaneously vary all four parameters, \dela, $N$, $b$ and $z$. 
On the contrary, we minimize \chisq by varying $N$, $b$ and $z$ for
each value of \dela. We then determine the minima of the 
\chisq versus \dela curve. Note that the two
 methods are equivalent as shown by Press et
 al. (2000, p. 691). For the sake
 of completeness we give details in Appendix A.\par
\begin{figure*} 
\psfig{figure=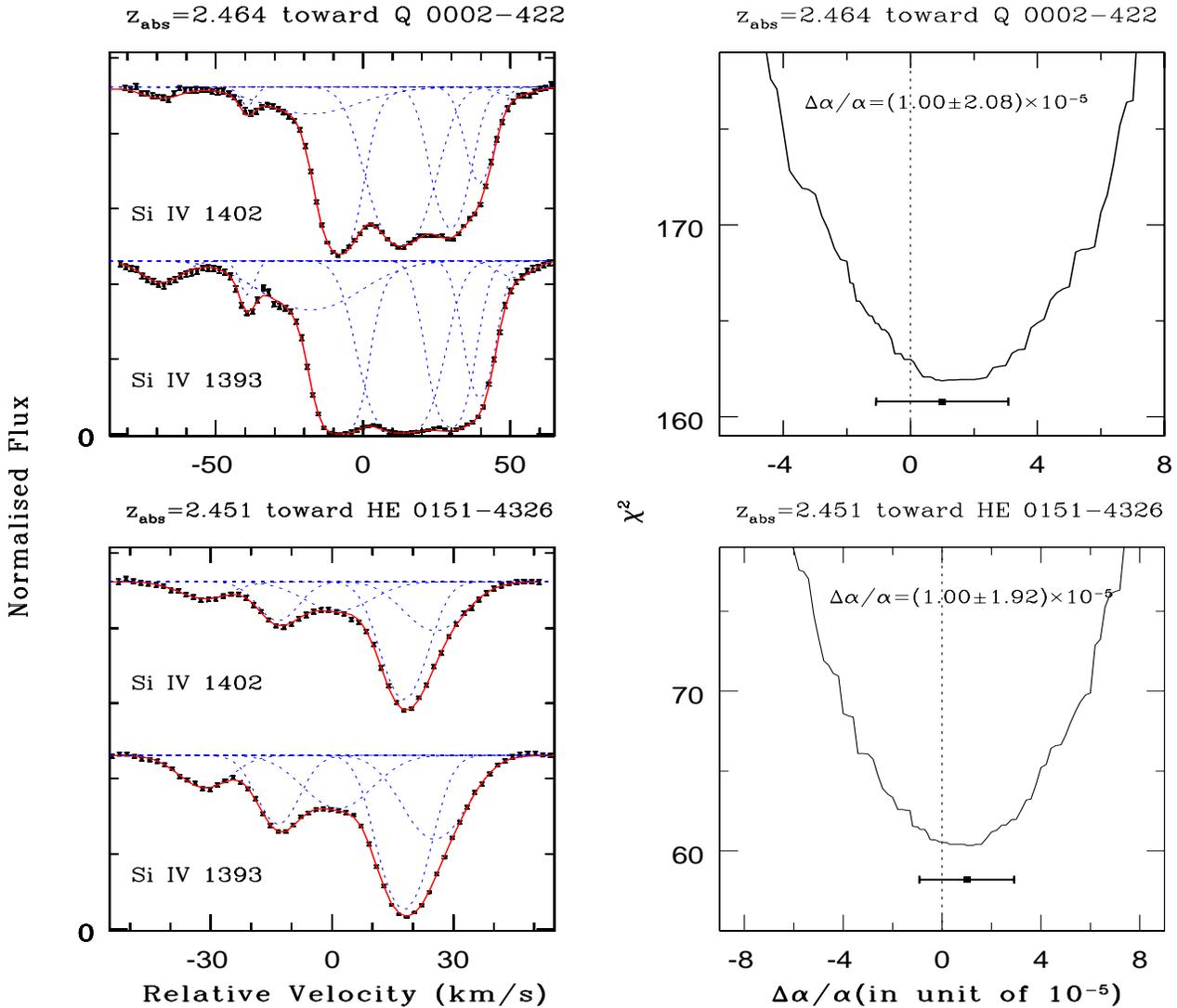,height=15.cm,width=18.cm,angle=0.0} 
\caption[]{Left panels shows on a velocity scale Si~{\sc iv}  doublet
data points with error-bars together with the best Voigt-profile fit  
for \dela$=0$ over plotted as a solid line. The Voigt profiles of individual  
 sub-components are shown by dashed lines. 
Right panels show the variation of $\chi^2$ 
as a function of \dela. The dark rectangle with error bars  
indicates the measured value of \dela with one sigma error-bars  
 derived  using the criteria $\Delta\chi^{2}=1$ around the minima. The name of QSOs,  
 \zabs~ and the value of measured \dela~ are also stated explicitly.}   
\label{2p462p45} 
\end{figure*}
\begin{figure*} 
\psfig{figure=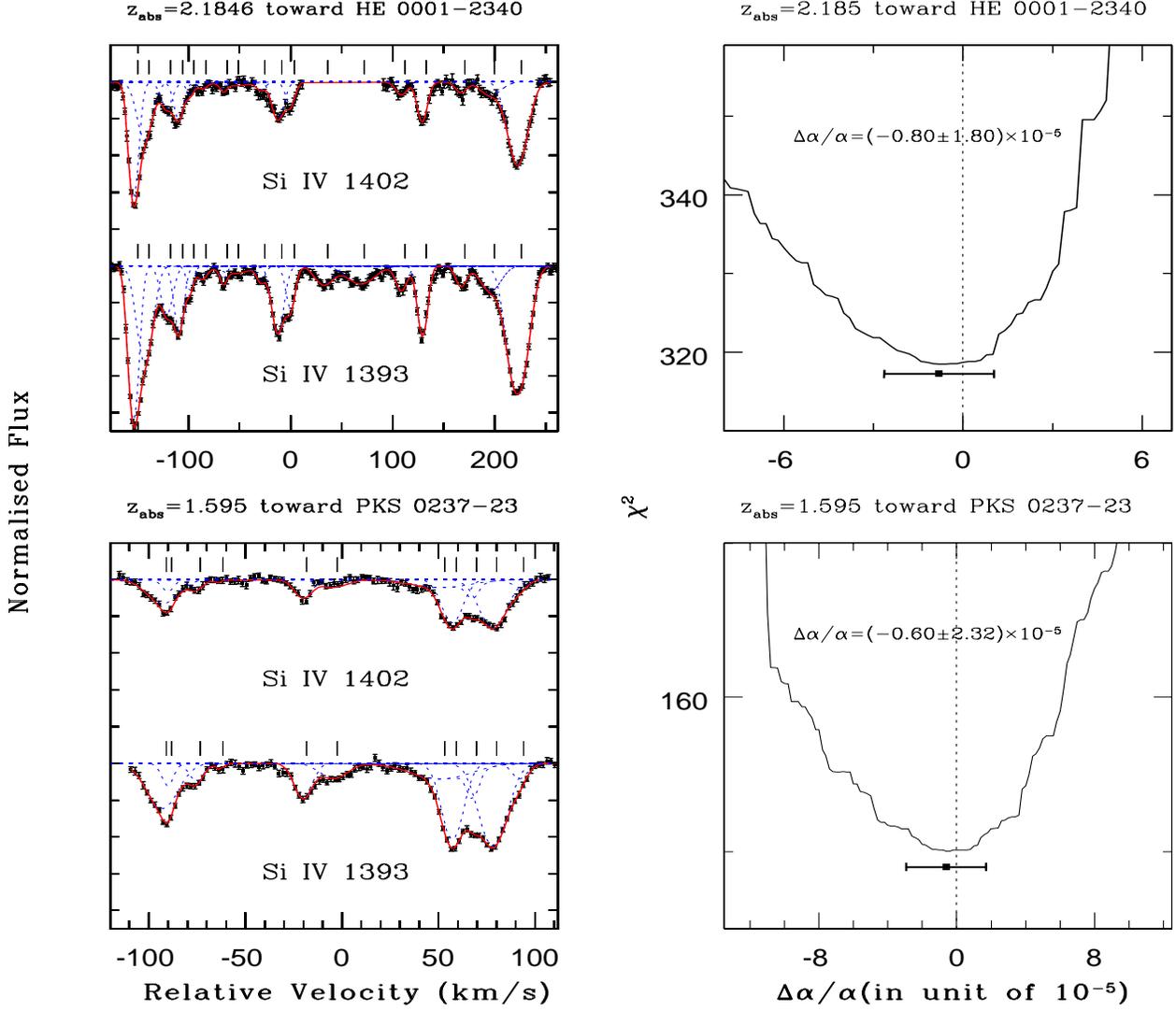,height=15.cm,width=18.cm,angle=0.0} 
\caption[]{Same as Fig.~\ref{2p462p45}. In addition, the position 
 of sub-components is also marked by tick marks. 
} 
\label{2p181p59} 
\end{figure*}

The Voigt profile fits to 11 individual systems 
and the corresponding plots giving $\chi^2$ as a function of \dela are  
presented in Fig.~\ref{fig1fit} and \ref{fig2fit}.
The Voigt profile fits to the other four systems, 
resulting in much more precise measurements of \dela, are presented
in more detail in Fig.~\ref{2p462p45} and  Fig.~\ref{2p181p59} (see Sect. 4). 
The $\chi^2$ curves 
presented for all the systems 
are median smoothed over \dela$=0.9\times 10^{-5}$, to avoid the 
obvious local fluctuations caused by one or two points in the
 curve. Since the smoothing scale we use is about one-half of the 
 limiting accuracy from individual Si~{\sc iv} doublets, the
 final result is not affected by the procedure.  
The results of the Voigt profile fits are summarized in Table~\ref{tabdet}. 
In this table, Cols. 1 and 2 give the QSOs name and the emission redshift ($z_{\rm em}$). 
Cols. 3, 4 and 5 give, respectively, the mean absorption redshift (\={z}$_{\rm abs}$) of 
all the components in the system, the measured \dela value and the reduced $\chi^{2}$ of 
the best-fit. The description regarding the component structure is provided in 
columns 6 to 9. 
In column 6 gives \zabs~ for individual components, while Cols. 7 and 8  
respectively list the column density and velocity dispersion.
Column 9 refers to the velocity of individual  
components relative to \={z}$_{\rm abs}$ (listed in column 3).
\par
In the course of the analysis we also noticed a few systems having weak components
on the edge of stronger components (a specific example is discussed in Sect. 4.1).
The parameters ($N$, $b$, $z$) of the weak components are difficult to constrain from
an overall fit. We therefore froze alternatively one of the
parameters (among $N$, $b$ and $z$) of the weak component and found that the 
final \dela 
does not depend much on the choice of the frozen parameter. To be on the conservative 
side we have taken as the final error the largest error of all determinations. 
Such systems have zero errors for the corresponding frozen parameter in Table~3. 
\par\noindent

\section{Importance of the selection}

\subsection{Errors versus strength of the absorption}
As an illustration of the importance of the selection of the systems, 
we plot in Fig.~\ref{figcri} \dela measured 
from the 20 Si~{\sc iv} systems in our sample versus the column density of the 
strongest component in the system. Open circles are for systems that we
define as "blended" (they do not pass our selection criteria) and filled 
circles are for systems that pass this selection criteria. 
It is apparent that errors are larger for weaker systems and for blended systems
as expected from the simulations presented in Paper I.
The column density of the strongest 
component is considered rather than the total column density since the  precision in   
\dela~measurements is most often dominated by the strongest component. 
\subsection{Precise measurement versus local minima} 
We illustrate here the importance of the selection criteria with an example 
of a spurious precise measurement which results from  
local minima caused by an unstable fit. This happens in the  
rejected system at \zabs = 2.8347 system toward HE~0940$-$1050. 
The Voigt profile best-fit (along with the profile of sub-components) and  
the variation of $\chi^{2}$ as a function of \dela for this system are 
shown, respectively, in the left and right panels of Fig.~\ref{fig2p83}. 
The system is fitted with 5 components and  $\chi^2_{\nu}=0.97$. 
The  right panel of Fig.~\ref{fig2p83} shows that  
the curve giving $\chi^{2}$  for this system possesses a large local fluctuation. 
 The cause of such fluctuation is that the fit is unstable, either due to the
uncertainty in the overall component structure or to abnormal fluctuation
in a few pixels (see Fig.~\ref{fig2p83} around $v=17$~\kms\ in \siivb).
The uncertainty on the measurement as derived from this curve 
using our procedure would be underestimated.
This kind of measurement could eventually dominate the whole statistics of 
weighted mean in the final result.  It is important to mention 
  that if we had not used the \chisq versus \dela method, we
  may have considered this local minimum as a true minimum (Fig.~\ref{fig2p83}).
\section{Notes on individual systems}
\begin{figure*} 
\centerline{\vbox{ 
\psfig{figure=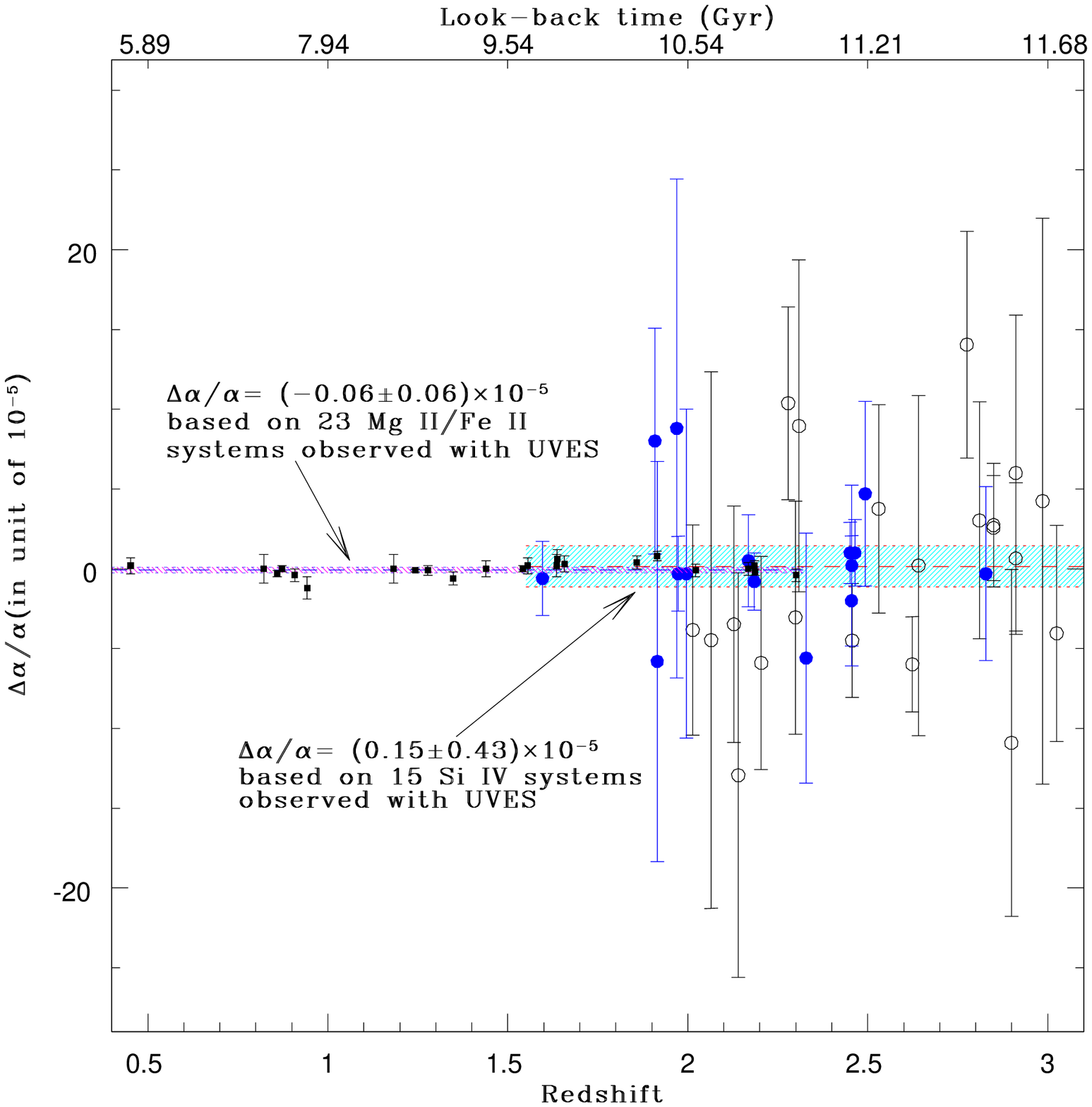,height=14.cm,width=18.cm,angle=0.} 
}} 
\caption[]{
Measured values of $\Delta\alpha/\alpha$ versus the absorption redshift of the systems. 
The squares show  our determinations from the MM method using
Mg~{\sc ii}/ Fe~{\sc ii} systems 
and the narrow shaded region represents the 3$\sigma$ allowed range. 
The filled circles are our present measurements using Si~{\sc iv} doublets from UVES 
and  open  circles are the measurements from KECK/HIRES data by Murphy et al. (2001).  
The weighted mean  from our 15 systems is  
${ \Delta\alpha/\alpha}$~=~${ (0.15\pm0.43)\times10^{-5}}$.
The $3\sigma$ allowed range   
($-1.14\times10^{-5}\le \Delta\alpha/\alpha \le 1.44\times 10^{-5}$) 
is shown by the wider shaded region.  
} 
\label{figfinal} 
\end{figure*} 
\begin{table*} 
\caption{Results of the Voigt profile fits to the 15 Si~{\sc iv} systems
         those satisfy our selection criteria.} 
{\tiny 
\begin{tabular}{lllllllrr} 
\hline 
\hline 
\\
\multicolumn{1}{c}{Name}&\multicolumn{1}{c}{$z_{\rm em}$}& \multicolumn{1}{c}{\={z}$_{abs}$}&\multicolumn{1}{c}{$\Delta\alpha/\alpha$} &\multicolumn{1}{c}{$\chi^{2}_{\nu}$} &\multicolumn{4}{c}{Component Structure}  \\ 
    &            &\multicolumn{1}{c}{(mean)}&\multicolumn{1}{c}{(in units of $10^{-5}$)} &           &\multicolumn{1}{c}{\zabs} &\multicolumn{1}{c}{log$_{10}$(N) cm$^{-2}$} & \multicolumn{1}{c}{b(\kms)} & \multicolumn{1}{c}{V(\kms)}     \\ 
\\
\hline 
\\
HE~1341$-$1020& 2.135  & 1.91534& $-5.80\pm$12.54  & 0.44 &$1.915491\pm0.000800$  &$  12.447\pm  0.023$  &$   8.669\pm  0.632$  &$  15.530$\\ 
             &        &        &  &                      &$1.915210\pm0.003547$  &$  11.799\pm  0.145$  &$   5.279\pm  2.652$  &$ -13.383$\\ 
             &        &        &  &                      &$1.915319\pm0.002501$  &$  11.751\pm  0.162$  &$   3.333\pm  2.404$  &$  -2.171$\\ 
\\
Q~0122$-$380  & 2.190  & 1.90870  &$ 8.00\pm7.08$  & 0.80 &$1.908656\pm0.000725$  &$  12.597\pm  0.027$  &$   2.651\pm  0.561$  &$  -4.586$\\ 
             &        &        &  &                      &$1.908745\pm0.000940$  &$  12.400\pm  0.042$  &$   2.863\pm  0.867$  &$   4.598$\\ 
             &        & 1.96961&$ 8.80\pm15.64$ & 0.52   &$1.969496\pm0.000963$  &$  12.364\pm  0.026$  &$   8.185\pm  0.715$  &$ -11.766$\\ 
             &        &        &  &                      &$1.969729\pm0.003927$  &$  11.851\pm  0.098$  &$   9.433\pm  2.975$  &$  11.766$\\ 
             &        & 1.97337&$-0.30\pm2.34$  & 0.87   &$1.973533\pm0.001488$  &$  13.188\pm  0.028$  &$  11.869\pm  0.000$  &$  16.779$\\
             &        &        &  &                      &$1.973729\pm0.000358$  &$  12.547\pm  0.021$  &$   3.191\pm  0.326$  &$  36.552$\\ 
             &        &        &  &                      &$1.973096\pm0.001843$  &$  12.730\pm  0.050$  &$   7.905\pm  0.873$  &$ -27.315$\\ 
             &        &        &  &                      &$1.973275\pm0.001371$  &$  13.207\pm  0.186$  &$   4.122\pm  1.780$  &$  -9.261$\\ 
             &        &        &  &                      &$1.973192\pm0.001700$  &$  12.922\pm  0.114$  &$   2.735\pm  0.841$  &$ -17.633$\\ 
             &        &        &  &                      &$1.973377\pm0.002469$  &$  13.543\pm  0.083$  &$   8.025\pm  1.193$  &$   1.034$\\ 
             &        & 1.97587&$-0.30\pm10.30$ & 0.58   &$1.975917\pm0.002038$  &$  12.303\pm  0.214$  &$   4.522\pm  1.177$  &$   4.711$\\ 
             &        &        &  &                      &$1.975823\pm0.012910$  &$  12.064\pm  0.374$  &$   8.132\pm  5.132$  &$  -4.759$\\
PKS~0237$-$23   &  2.223 & 1.59671&$-0.60\pm2.32$  & 0.82   &$1.595925\pm0.001055$  &$  11.763\pm  0.098$  &$   1.160\pm  2.414$  &$ -90.264$\\ 
             &        &        &  &                      &$1.595902\pm0.001960$  &$  12.464\pm  0.060$  &$  10.260\pm  0.851$  &$ -92.922$\\ 
             &        &        &  &                      &$1.596053\pm0.001286$  &$  11.703\pm  0.112$  &$   3.052\pm  1.549$  &$ -75.473$\\ 
             &        &        &  &                      &$1.597202\pm0.000911$  &$  12.582\pm  0.159$  &$   6.297\pm  1.082$  &$  57.265$\\ 
             &        &        &  &                      &$1.597384\pm0.001517$  &$  12.736\pm  0.136$  &$   8.762\pm  1.948$  &$  78.296$\\ 
             &        &        &  &                      &$1.597151\pm0.048581$  &$  12.253\pm  0.704$  &$  21.079\pm 15.835$  &$  51.385$\\ 
             &        &        &  &                      &$1.597293\pm0.001307$  &$  11.873\pm  0.256$  &$   1.650\pm  2.934$  &$  67.788$\\ 
             &        &        &  &                      &$1.597504\pm0.003343$  &$  11.676\pm  0.357$  &$   3.785\pm  2.695$  &$  92.165$\\ 
             &        &        &  &                      &$1.596531\pm0.000759$  &$  12.104\pm  0.041$  &$   4.919\pm  0.704$  &$ -20.245$\\ 
             &        &        &  &                      &$1.596669\pm0.002426$  &$  11.898\pm  0.077$  &$   8.786\pm  2.127$  &$  -4.311$\\ 
             &        &        &  &                      &$1.596154\pm0.000337$  &$  11.013\pm  0.202$  &$   0.276\pm  0.128$  &$ -63.807$\\ 
\\
HE~0001$-$2340& 2.263  & 2.18469&$-0.80\pm1.80$  & 1.14   &$2.183055\pm0.000661$  &$  13.121\pm  0.063$  &$   4.817\pm  0.352$  &$-153.733$\\ 
             &        &        &  &                      &$2.183171\pm0.004262$  &$  12.855\pm  0.133$  &$   9.143\pm  2.517$  &$-142.796$\\ 
             &        &        &  &                      &$2.183524\pm0.002198$  &$  12.453\pm  0.184$  &$   5.042\pm  2.232$  &$-109.556$\\ 
             &        &        &  &                      &$2.183398\pm0.004220$  &$  12.367\pm  0.224$  &$   6.609\pm  3.394$  &$-121.415$\\ 
             &        &        &  &                      &$2.183642\pm0.002869$  &$  11.853\pm  0.231$  &$   2.758\pm  3.389$  &$ -98.439$\\ 
             &        &        &  &                      &$2.183769\pm0.003596$  &$  11.654\pm  0.212$  &$   4.851\pm  3.711$  &$ -86.468$\\ 
             &        &        &  &                      &$2.184555\pm0.001184$  &$  12.534\pm  0.046$  &$   6.542\pm  0.987$  &$ -12.420$\\ 
             &        &        &  &                      &$2.184685\pm0.001782$  &$  12.204\pm  0.084$  &$   4.511\pm  0.993$  &$  -0.180$\\ 
             &        &        &  &                      &$2.184380\pm0.002295$  &$  11.676\pm  0.147$  &$   3.821\pm  2.283$  &$ -28.905$\\ 
             &        &        &  &                      &$2.186058\pm0.000353$  &$  12.502\pm  0.012$  &$   5.220\pm  0.268$  &$ 129.163$\\ 
             &        &        &  &                      &$2.185840\pm0.001094$  &$  12.021\pm  0.035$  &$   6.257\pm  0.893$  &$ 108.613$\\ 
             &        &        &  &                      &$2.187051\pm0.000462$  &$  13.158\pm  0.008$  &$  12.018\pm  0.230$  &$ 222.706$\\ 
             &        &        &  &                      &$2.186770\pm0.003257$  &$  12.290\pm  0.062$  &$  13.227\pm  2.068$  &$ 196.226$\\ 
             &        &        &  &                      &$2.186462\pm0.001357$  &$  11.922\pm  0.044$  &$   5.964\pm  0.965$  &$ 167.209$\\ 
             &        &        &  &                      &$2.183988\pm0.002104$  &$  11.575\pm  0.219$  &$   0.739\pm  1.589$  &$ -65.828$\\ 
             &        &        &  &                      &$2.185035\pm0.002420$  &$  12.100\pm  0.052$  &$  11.632\pm  1.599$  &$  32.790$\\ 
             &        &        &  &                      &$2.185417\pm0.003240$  &$  12.276\pm  0.044$  &$  18.832\pm  2.403$  &$  68.770$\\ 
             &        &        &  &                      &$2.184102\pm0.011745$  &$  11.914\pm  0.148$  &$  20.314\pm  7.976$  &$ -55.092$\\ 
\\
HE~1347$-$2457& 2.611 &  2.32918&$-5.60\pm7.84$  & 0.98 &$2.328624\pm0.001374$  &$  12.286\pm  0.029$  &$  10.056\pm  0.985$  &$ -50.424$\\ 
             &        &        &  &                      &$2.328948\pm0.000556$  &$  12.616\pm  0.014$  &$   8.712\pm  0.410$  &$ -21.227$\\ 
             &        &        &  &                      &$2.329217\pm0.001300$  &$  12.111\pm  0.040$  &$   7.045\pm  0.970$  &$   3.008$\\ 
             &        &        &  &                      &$2.329599\pm0.003715$  &$  12.338\pm  0.058$  &$  16.509\pm  1.703$  &$  37.426$\\ 
             &        &        &  &                      &$2.329531\pm0.000592$  &$  12.281\pm  0.050$  &$   3.913\pm  0.611$  &$  31.303$\\ 
\\
PKS~0329$-$255& 2.685  & 2.45465&$-2.00\pm4.07$  & 0.47   &$2.454707\pm0.001053$  &$  12.579\pm  0.149$  &$   3.141\pm  1.277$  &$   4.886$\\ 
             &        &        &  &                      &$2.454874\pm0.001173$  &$  12.542\pm  0.062$  &$   4.178\pm  0.598$  &$  19.400$\\ 
             &        &        &  &                      &$2.454643\pm0.002560$  &$  13.215\pm  0.056$  &$  11.463\pm  0.896$  &$  -0.663$\\ 
             &        &        &  &                      &$2.454378\pm0.004649$  &$  12.868\pm  0.065$  &$  16.186\pm  1.530$  &$ -23.686$\\ 
             &        & 2.45573&$0.20\pm5.03$   & 0.36   &$2.455683\pm0.001245$  &$  12.593\pm  0.177$  &$   3.340\pm  0.913$  &$  -3.705$\\ 
             &        &        &  &                      &$2.455768\pm0.007809$  &$  12.496\pm  0.236$  &$   7.026\pm  2.405$  &$   3.684$\\ 
\\
Q~0002$-$422 &  2.760  & 2.16816&$ 0.50\pm2.89$  & 1.33   &$2.167774\pm0.000928$  &$  12.509\pm  0.028$  &$   6.464\pm  0.000$  &$ -36.100$\\ 
             &        &        &  &                      &$2.167906\pm0.000245$  &$  13.265\pm  0.007$  &$   6.198\pm  0.128$  &$ -23.592$\\ 
             &        &        &  &                      &$2.168234\pm0.000308$  &$  13.430\pm  0.006$  &$   8.496\pm  0.164$  &$   7.473$\\ 
             &        &        &  &                      &$2.168391\pm0.000357$  &$  12.985\pm  0.026$  &$   4.172\pm  0.327$  &$  22.328$\\ 
             &        &        &  &                      &$2.168473\pm0.001424$  &$  12.750\pm  0.039$  &$   9.784\pm  0.000$  &$  30.094$\\ 
             &        & 2.46401&$1.00\pm2.08$   & 1.13   &$2.463220\pm0.001224$  &$  12.068\pm  0.031$  &$   8.850\pm  0.783$  &$ -68.511$\\ 
             &        &        &  &                      &$2.463558\pm0.000494$  &$  11.898\pm  0.037$  &$   1.515\pm  0.791$  &$ -39.232$\\ 
             &        &        &  &                      &$2.464161\pm0.001020$  &$  13.713\pm  0.027$  &$  10.864\pm  0.673$  &$  12.988$\\ 
             &        &        &  &                      &$2.464356\pm0.001321$  &$  13.382\pm  0.092$  &$   6.834\pm  1.105$  &$  29.878$\\ 
             &        &        &  &                      &$2.464470\pm0.002380$  &$  12.977\pm  0.142$  &$   5.334\pm  0.553$  &$  39.748$\\ 
             &        &        &  &                      &$2.463805\pm0.016700$  &$  12.848\pm  0.160$  &$  22.392\pm  4.165$  &$ -17.840$\\ 
             &        &        &  &                      &$2.463918\pm0.000331$  &$  13.651\pm  0.016$  &$   6.757\pm  0.194$  &$  -8.053$\\ 
             &        &        &  &                      &$2.464605\pm0.000000$  &$  11.885\pm  0.049$  &$   7.291\pm  1.068$  &$  51.455$\\ 
\\
HE~0151$-$4326& 2.740  & 2.45140&$1.00\pm1.92$   & 0.54   &$2.451609\pm0.000305$  &$  13.190\pm  0.026$  &$   5.991\pm  0.231$  &$  17.802$\\ 
             &        &        &  &                      &$2.451421\pm0.001505$  &$  12.585\pm  0.113$  &$  10.638\pm  3.157$  &$   1.471$\\ 
             &        &        &  &                      &$2.451696\pm0.001122$  &$  12.797\pm  0.039$  &$   9.311\pm  0.000$  &$  25.366$\\ 
             &        &        &  &                      &$2.451250\pm0.000908$  &$  12.519\pm  0.074$  &$   5.674\pm  0.494$  &$ -13.387$\\ 
             &        &        &  &                      &$2.451043\pm0.000683$  &$  12.228\pm  0.017$  &$   7.899\pm  0.444$  &$ -31.396$\\
             &        & 2.49265&$4.70\pm5.79$   & 0.60   &$2.492705\pm0.000453$  &$  12.297\pm  0.015$  &$   4.820\pm  0.297$  &$   4.894$\\ 
             &        &        &  &                      &$2.492591\pm0.002372$  &$  11.400\pm  0.093$  &$   3.261\pm  0.000$  &$  -4.915$\\ 
\\
HE~0940$-$1050& 3.084  & 2.82831&$-0.30\pm5.46$  & 0.68   &$2.827964\pm0.003259$  &$  12.413\pm  0.087$  &$   7.757\pm  1.541$  &$ -26.941$\\ 
             &        &        &  &                      &$2.828107\pm0.000997$  &$  12.501\pm  0.070$  &$   4.463\pm  0.618$  &$ -15.731$\\ 
             &        &        &  &                      &$2.828288\pm0.002861$  &$  12.493\pm  0.154$  &$   4.490\pm  1.076$  &$  -1.551$\\ 
             &        &        &  &                      &$2.828393\pm0.006348$  &$  12.267\pm  0.235$  &$   5.232\pm  1.842$  &$   6.670$\\ 
             &        &        &  &                      &$2.828776\pm0.005376$  &$  12.447\pm  0.068$  &$  15.967\pm  1.289$  &$  36.675$\\ 
             &        &        &  &                      &$2.828899\pm0.000903$  &$  12.369\pm  0.078$  &$   6.716\pm  0.738$  &$  46.316$\\ 
             &        &        &  &                      &$2.827730\pm0.001076$  &$  11.990\pm  0.040$  &$   4.449\pm  0.712$  &$ -45.288$\\ 

\\
\hline 
\end{tabular} 
\label{tabdet} 
} 
\end{table*} 
Table~\ref{tabdet} shows that some of  
our measurements have error-bars on \dela~ less than or  
comparable to $2\times10^{-5}$ which is much smaller than the average.  
Such systems need to be discussed in more detail because 
they will dominate the weighted mean of the \dela measurements.  
\subsection{\zabs = 2.464 system toward Q~0002$-$422} 
This system is spread over a velocity range of about 240 \kms. 
Here we have used only the well detached red part of the system, as
the profile of the \siiva line in the blue part is 
affected by a bad pixel.
The best-fit Voigt profile  and the profiles of 
its sub-components are shown in the top left panel of Fig.~\ref{2p462p45}. 
As is evident from the figure this system is well fitted with eight  
sub-components ($\chi^2_{\nu}=1.13$). 
The red-most component of this system (V$\approx 52$~\kms) 
is found to make the Voigt profile fit of the system unstable, if we vary all 
its parameters at the same time. 
As a result we have frozen one of the parameters to their best-fit
value obtained assuming \dela=0. We performed \dela measurements for the three
cases of freezing one of its parameter among $N$, $b$, and $z$ 
(as discussed in Sect. 2.2). We find that all  
measurements are consistent with one another and accept the 
measurements with largest error (i.e \dela = $(1.00\pm2.08)\times10^{-5}$).
The relatively stronger constraint is mainly due to  
the presence of 3 well separated strong components together with  
good S/N ratios ($\sim68$ per pixel in the nearby continuum). 
\subsection{\zabs = 2.451 system toward HE~0151$-$4326} 
The best-fit Voigt profiles and variation of $\chi^{2}$ as a function of  
\dela for this system are shown in the bottom panels of Fig.~\ref{2p462p45}. 
The system is very well fitted by 5 components with $\chi^2_{\nu}=0.54$.  
The measured \dela value from this system is $(1.00\pm1.92)\times10^{-5}$. 
The good accuracy is mainly due to 3 well separated  
strong components and very good S/N ($\sim89$ per pixel in the nearby continuum). 
\subsection{\zabs = 2.1839 system toward HE~0001$-$2340} 
The top panels of Fig.~\ref{2p181p59} show the best-fit Voigt profiles  
and variation of $\chi^{2}$ as a function of \dela for this system. 
The absorption profiles produced by this system are spread over 400 \kms, 
but most of the components are very well separated. The system is fitted 
by 18 sub-components (indicated by tick-marks) 
with $\chi^2_{\nu}=1.14$. We have excluded from the fit the velocity range 
32 to 68 \kms~ in the \siivb profile because of the presence of several spurious 
pixels. 
The measured \dela value from this system is 
$(-0.80\pm1.80)\times10^{-5}$. The S/N ratio of the spectrum in the 
vicinity of this system is about 52. The presence of a large number of  
unblended components many of which are narrow and strong increases 
the precision of the \dela measurement.  
\begin{table*} 
\caption{ {\bf Summary of results from UVES and HIRES samples:}} 
\centerline{  
\begin{tabular}{lccccrr} 
\hline 
\hline 
\multicolumn {1}{c}{Sample}& Numbers & \multicolumn{1}{c}{z}& \multicolumn{3}{c}{\dela($10^{-5}$)}&$\chi^2_{\rm w}$\\ 
                           & of systems&\multicolumn{1}{c}{range}&\multicolumn {1}{c}{weighted mean} & \multicolumn {1}{c}{mean}&\multicolumn{1}{c}{$\sigma$}&\\ 
\hline 
UVES      & 15 & 1.59$-$2.82 &$+0.15\pm0.43   $&$+0.57\pm1.05 $&4.06  &0.29\\ 
HIRES        & 21 & 2.01$-$3.02 &$-0.52\pm1.22      $&$-0.12\pm1.48   $&6.80  &0.95\\ 
UVES+HIRES   &  36& 1.59$-$3.02 &$-0.04\pm0.56  $&$+0.17\pm0.96 $&5.76 &0.67 \\ 
\hline 
\end{tabular} } 
\label{tabres}                                     
\end{table*} 
\subsection{\zabs = 1.59671 system toward PKS~0237$-$23} 
This system falls in the Lyman-$\alpha$ forest but it has 
an unblended structure with well separated components; as a result we have included it in  
our analysis.  The absorption profile of Si~{\sc iv} in this system is spread over  
about 240 \kms. The best-fit Voigt profile along with profile of  
the different sub-components are shown in the  bottom left panel of  
Fig.~\ref{2p181p59}. The system is fitted with 11 components with 
$\chi^2_{\nu}=0.82$. The average S/N ratio per pixel in the neighboring continuum 
is about 57. The measured \dela~ value from this 
system is $(-0.60\pm2.32)\times10^{-5}$. The relatively good 
precision 
is mainly due to the presence of a large number of well separated  
and strong enough sub-components. 
\section{Results and discussion} 
The detailed description of our individual measurements from  
15 Si~{\sc iv} doublets is given in
Table~\ref{tabdet}. The summary of the results 
obtained from different samples is presented in Table~\ref{tabres}. 
In column 1 ``UVES'' refers to our sample
 and  ``HIRES'' refers to the KECK/HIRES sample from Murphy et al. (2001). 
Cols. 2 and 3 list, respectively, the number of systems in the sample and  
their redshift coverage. The weighted mean and mean value of 
 \dela are listed respectively in Cols. 4 and 5. Cols. 6, 7 give 
 the standard deviation of measurements around the mean and the reduced $\chi^{2}$ 
 around the weighted mean. \\ 
 The weighted mean of the  measurements is obtained 
 by assigning weights ($w_{\rm i}$) as 1/error$^{2}$ and 
 the error on the weighted mean is computed by the standard equation, 
\begin{equation} 
{\rm 
Error ~in}~x_{\rm w} ~=~ \sqrt{\chi^{2}_{w}\over  
{{\rm \Sigma_i^N}w_{\rm i} }}. 
\end{equation} 
Here $\chi^{2}_{w}$ refers to the reduced $\chi^{2}$ of variable $x_{\rm i}$  
around their weighted mean $x_{\rm w}$.
The $\chi^{2}_{w}$ term takes into account the 
effect of scatter in measurements while computing the error on the weighted mean \dela.
 The error on the simple mean \dela ( column 5 of Table~\ref{tabres}) is 
 computed by the central limit theorem ($\sigma/\sqrt N$ ), assuming the individual measurements are 
 Gaussian distributed around their mean.\par  
The distribution of our 15 measurements together with the 21 measurements of
Murphy et al. (2001) is plotted  as a function of \zabs~ and look-back time 
in Fig.~\ref{figfinal}. The look-back time corresponding to a given 
redshift is computed in the case of a flat universe with 
$\Omega_\lambda = 0.7$, $\Omega_{\rm m}$ = 0.3 and $H_{\rm 0}$ = 68 \kms Mpc$^{-1}$.   
The measurements from our UVES sample using the MM method on Mg~{\sc ii} systems 
(Paper I) are also shown for comparison. 
The weighted mean value obtained from our analysis 
over the redshift range ${ 1.59\le z\le 2.82}$ is  
${ \Delta\alpha/\alpha}$~=~${ (0.15\pm0.43)\times10^{-5}}$. 
The $3\sigma$ range   
($-1.14\times10^{-5}\le \Delta\alpha/\alpha \le 1.44\times 10^{-5}$) 
is shown in Fig.~\ref{figfinal} as a shaded region. 
Our result corresponds to a factor of three improvement on the constraint based on \siiv 
doublets compared to the previous study by Murphy et al. (2001). 
The increased accuracy in our result is mainly due to 
the better quality of the data (S/N ratio of about $\sim~70$ per pixel and $R\sim45000$). 
Combining our sample with  KECK/HIRES sample 
results in a weighted mean ${ \Delta\alpha/\alpha}$~=~${ (-0.04\pm0.56)\times10^{-5}}$ 
over a redshift range of $1.59< z< 3.02$. The small enhancement in the 
error in the weighted mean \dela in the case of the combined sample is due to higher 
$\chi^{2}_{w}$.  
\par
Further improvements at higher redshift can be achieved using
 as MM analysis of multiplets from 
single species such as Ni~{\sc ii} or Fe~{\sc ii} using a well defined high
 quality sample (see e.g. Quast et al. 2004). It is also demonstrated that OH and other 
molecular lines can be used to improve limits on the variation of $\alpha$ 
(see for example Chengalur \& Kanekar, 2003). 
In addition other constants can be constrained in a similar way.
Although it is hard to make any quantitative prediction 
theorists estimate that variations in the proton-to-electron mass ratio 
could be larger than that of the fine-structure constant by  a factor of 10 to 50. 
It  is possible to constrain this constant by 
measuring the wavelengths of radiative transitions produced by
molecular hydrogen, H$_{2}$. On-going ESO programes have been dedicated
to this purpose (Ledoux et al. 2003, Ivanchik et al. 2002, Petitjean et al. 2004b).  
\section*{Acknowledgments} 
This work is based on observations collected during programme 166.A-0106 
(PI: Jacqueline Bergeron) of the European Southern Observatory with the 
Ultra-violet and Visible Echelle 
Spectrograph mounted on the 8.2~m Kuyen telescope operated at the Paranal 
Observatory, Chile. PPJ thanks E. Vangioni-Flam and J. P. Uzan for 
fruitful discussions. 
HC thanks CSIR, INDIA for the grant award 
No. 9/545(18)/2KI/EMR-I and CNRS/IAP for the hospitality. 
We gratefully acknowledge support from the Indo-French 
Centre for the Promotion of Advanced Research (Centre Franco-Indien pour 
la Promotion de la Recherche Avanc\'ee) under contract No. 3004-3.  

\appendix
\label{errTwoMethod}

\section{Relation between errors
  $\delta(\Delta\alpha/\alpha)$ estimated from the \chisq versus \dela 
curve and the  covariance matrix $\sqrt{C_{11}}$} 
In general to fit a nonlinear function like a Voigt-profile, one would
define the merit function \chisq
 \[\chi^{2}(a)=
 \mathop{\sum_{j=1}^{N}}\bigg{[}\frac{(y_{j}-y(x_{j}:a))^2}{\sigma_{j}^2}\bigg{]} \]    
and minimize it to get the best-fit value of the parameters
$a_{k}$ where k=1, 2, ... M. The minimization of such a nonlinear function
involves an iterative process. In a  given iteration the trial values of
parameters are improved till one reaches \chisq minima.\par
Let us suppose that with sufficient accuracy, we can express
$\chi^{2}(a)$ as a quadratic function near its minimum. Then
\begin{equation}
\chi^{2}(a)~\approx~\gamma-d\cdot a +\frac{1}{2}a\cdot D\cdot a 
\end{equation}
where $``d"$  is a row matrix of M elements and $``D"$ 
is a $ M\times M $ matrix (Press et al. 2000, p. 675).\par
For such a quadratic function to go from the current parameter $a_{cur}$
to $a_{min}$, which is the parameter at $\chi^{2}_{min}$, one can use
a Newton method of minimization (almost equivalent
to a variable metric or Hessian matrix method).
It states that if $a_{min}$ is the parameter at $\chi^{2}_{min}$ then
$\nabla\chi^{2}(a_{min})=0$.\par
Now let us suppose  $a_{cur}$ is close enough to $a_{min}$ so that 
to second order we can write
\begin{eqnarray}
\chi^{2}(a_{min})&=&\chi^{2}(a_{cur})-(a_{min}-a_{cur})\cdot d \nonumber \\
                 & &+\frac{1}{2}(a_{min}-a_{cur})\cdot D\cdot(a_{min}-a_{cur})
\end{eqnarray}
where $``d"$ and $``D"$ correspond to the first and second
derivative terms evaluated at  $a_{cur}$.\\
The requirement that $\nabla\chi^{2}(a_{min})=0$ in Eq.(A.2) 
will give us
\begin{equation}
 D.\delta a=d
\end{equation}
where $\delta a =a_{min}-a_{cur}$ and \\ 
$$
 \begin{array}{rcll}
 D_{kl}=\frac{\partial^{2}\chi^{2}(a_{cur})}{\partial a_{l}a_{k}}, &
 d_{k}=-\frac{\partial \chi^{2}(a_{cur})}  {\partial a_{k}} 
\end{array}
$$
In a more familiar form Eq.(A.3) appears as 
\begin{equation}
\alpha_{kl}\cdot \delta a_{l}=\beta_{k}
\end{equation}
 where $\beta_{k}=\frac{1}{2}d_{k}$,  $\alpha_{kl}=\frac{1}{2}D_{kl}$
 and summation is assumed over the repeated indices. Now the whole
 effort of minimization is to make $\beta_{k}$ vanish in
 Eq.(A.4). At this points  $\delta a=0 $ and hence $a_{cur}$ will be
 equal to $a_{min}$.\par
Now we apply this general approach to our specific case of
four parameters   \dela, $N$, $b$ and $z$. Let us suppose that
$a_{min}=[a^{min}_{1},a^{min}_{2},a^{min}_{3},a^{min}_{4}]$ are  the
best-fit value of these parameters achieved by varying all four parameters and let
$\chi^{2}_{min}(a_{min})$ be the minimum value of \chisq. This
parameter set is not different from  the one that is obtained from 
the minima of the \chisq versus \dela curve. This is because the  \chisq
versus \dela curve  is obtained by standard \chisq minimization, 
by varying $N$, $b$ and $z$, at every value of \dela.
The reason we have chosen this method
is that the \chisq versus \dela curve will allow us to avoid the local minima.\par 
Now let us keep fixed the first element of our parameter vector 
with a value close to $a^{min}_{1}$, and let us vary the other three
parameters to achieve the \chisq minimization. The resulting parameter
set is $a_{c}=[a^{c}_{1},~a^{c}_{2},~a^{c}_{3},~a^{c}_{4}]$ ($a^{c}_1$
has been kept fixed) and the minimum value of \chisq is
$\chi^{2}_{min}(a_{c})$.\par
Now substituting $a_{cur}$ by $a_{c}$ in the more general equation Eq.(A.4), we get 
\begin{equation}
\delta a \cdot \alpha \cdot\delta a
=\Delta\chi^{2}=\chi^{2}_{min}(a_{c})- \chi^{2}_{min}(a_{min})
\end{equation}
\begin{equation}
\alpha \cdot \delta a =\beta
\end{equation}
where $\delta a=a_{min}-a_{c}$. But here we should keep in mind that
$a_{c}$ is also the best fit parameter obtained by \chisq minimization,
except that its first element was held fixed.
We know that for the best-fit parameters the corresponding elements of
$\beta$ should vanish. Therefore imposing all elements of 
$\beta$, except the first, to be zero, Eq.(A.6) becomes
\begin{equation}
 \alpha \cdot \delta a =\left( \begin{array}{c}
      c \\
      0 \\
      0 \\
      0 \end{array} \right)
\end{equation}
giving
\[\delta a_{1}/C_{11}=c \]
where we have used the fact that $\alpha$ is the inverse of
the covariance matrix $C$.
Combining Eq.(A.7) and Eq.(A.5) we get 
\[ (\delta a_{1})^{2} =\Delta\chi^{2} C_{11}\]
and using the fact that $C_{11}=\sigma_{1}^{2}$, where $\sigma_{1}$ is
the error-bar on the first parameter, we get the required relation
\begin{equation} 
\delta a_{1} =\pm \sqrt{\Delta\chi^{2}}\sigma_{1}. 
\end{equation} 
Interestingly $\delta a_{1}$ (the difference between two measurements of
first parameter) becomes equal to  $\sigma_{1}$
 (the error obtained using a covariance matrix, while varying all the
 four parameters) when $ \Delta\chi^{2}=1$.
Also when we use \chisq versus \dela curve we obtain the error in 
\dela using $\Delta\chi^{2}\simeq 1$. This means that the estimated errors
 using both the methods are identical. 
 
%
\end{document}